\documentclass[twocolumn,showpacs,aps,prd,nofootinbib]{revtex4}
\usepackage{graphicx}
\usepackage{dcolumn}
\usepackage{amsmath}
\usepackage{epsfig}
\usepackage{colordvi}
\usepackage{color}
\usepackage{hhline}
\usepackage{relsize}
\usepackage{hyperref}

\def\babar{\mbox{\slshape B\kern-0.1em{\smaller A}\kern-0.1em
    B\kern-0.1em{\smaller A\kern-0.2em R}}}
\def\pep2{PEP-II}

\begin{document}

\begin{flushleft}
\mbox{\normalsize  \babar\ -PUB-18/005} 

\mbox{\normalsize SLAC-PUB-17286}
\end{flushleft}
\vskip 20pt
\title{
 \large \bf \boldmath
Measurement of the spectral function for the 
$\tau^-\to K^-K_S\nu_{\tau}$ decay}
%
%
\author{J.~P.~Lees}
\author{V.~Poireau}
\author{V.~Tisserand}
\affiliation{Laboratoire d'Annecy-le-Vieux de Physique des Particules (LAPP), Universit\'e de Savoie, CNRS/IN2P3,  F-74941 Annecy-Le-Vieux, France}
\author{E.~Grauges}
\affiliation{Universitat de Barcelona, Facultat de Fisica, Departament ECM, E-08028 Barcelona, Spain }
\author{A.~Palano}
\affiliation{INFN Sezione di Bari and Dipartimento di Fisica, Universit\`a di Bari, I-70126 Bari, Italy }
\author{G.~Eigen}
\affiliation{University of Bergen, Institute of Physics, N-5007 Bergen, Norway }
\author{D.~N.~Brown}
\author{Yu.~G.~Kolomensky}
\affiliation{Lawrence Berkeley National Laboratory and University of California, Berkeley, California 94720, USA }
\author{M.~Fritsch}
\author{H.~Koch}
\author{T.~Schroeder}
\affiliation{Ruhr Universit\"at Bochum, Institut f\"ur Experimentalphysik 1, D-44780 Bochum, Germany }
% ...
\collaboration{The \babar\ Collaboration}
\noaffiliation

\begin{abstract}
The  decay $\tau^{-}\to K^{-}K_S\nu_{\tau}$ has been studied 
using $430\times10^6$ $e^+e^-\to \tau^+\tau^-$
events produced at a center-of-mass energy around 10.6 GeV
 at the PEP-II collider and studied with the  \babar\ detector.
The mass spectrum of the $K^{-}K_S$ system has been measured and the 
spectral function has been obtained. The 
measured branching fraction ${\cal B}(\tau^{-}\to K^{-}K_S\nu_{\tau})
= (0.739\pm 0.011(\rm stat.)\pm 0.020(\rm syst.))\times 10^{-3}$
is found to be in agreement with earlier measurements.
\end{abstract}


\maketitle

\section{Introduction}
The $\tau$ lepton provides a remarkable laboratory 
for studying many open questions in particle physics. 
With a large statistics of  about $10^9$ $\tau$'s produced in $e^+e^-$ 
annihilation at the \babar\ experiment, 
various aspects can be  studied,
for example, improving  the precision of 
spectral functions describing 
the mass distribution of  the hadronic decays of the  $\tau$.
In this work, we analyze 
the $\tau^{-}\to K^{-}K_S\nu_{\tau}$ decay\footnote{Throughout this paper,
inclusion of charge-conjugated channels  is implied.} 
and measure the spectral function of this channel defined as ~\cite{CVC}
\begin{equation}
V(q)=\frac{m_{\tau}^8}{12\pi C(q) |V_{ud}|^2}
\frac{{\cal B}(\tau^- \to K^-K_S\nu_\tau)}
{{\cal B}(\tau^- \to e^-\bar{\nu_e}\nu_\tau)}
\frac{1}{N}\frac{dN}{dq},
\label{eq1}
\end{equation}
where $m_{\tau}$ is the $\tau$ mass \cite{PDG},
$q\equiv m_{K^-K_S}$ is the invariant mass of the $K^-K_S$
system,  $V_{ud}$ is an 
element of the CKM matrix \cite{PDG}, 
$(dN/dq)/N$ is the normalized $K^-K_S$ mass 
spectrum,  and $C(q)$ is 
the phase space correction factor given by the following formula:
\begin{equation}
C(q)=q(m_{\tau}^2-q^2)^2(m_{\tau}^2+2q^2).
\label{eq2}
\end{equation}

 According to the  conserved-vector-current  hypothesis ~\cite {CVC},
the  $\tau^{-}\to K^{-}K_S\nu_{\tau}$ spectral function is related to
the isovector part ($I$=1) of the $e^+e^-\to K\bar{K}$ cross section: 
\begin{equation}
\sigma_{e^+e^-\to K\bar{K}}^{I=1}(q)=\frac{4\pi^{2}\alpha^{2}}{q^{2}}V(q),
\label{eq3}
\end{equation}
where $\alpha$ is the fine structure constant.
The cross sections $e^+e^-\to  K^+K^-$ and $e^+e^-\to  K_SK_{L}$ 
have been   recently measured 
by the \babar\ ~\cite{Kpkm,Kskl} and  SND experiments ~\cite{SndKK}. 
Combining data from
the  $\tau^{-}\to K^{-}K_S\nu_{\tau}$ with $e^+e^-\to  K\bar{K}$
measurements, the moduli of the isovector and isoscalar form factors and
the relative phase between them can obtained in a model-independent way.

\begin{figure}
\includegraphics[width=0.45\textwidth]{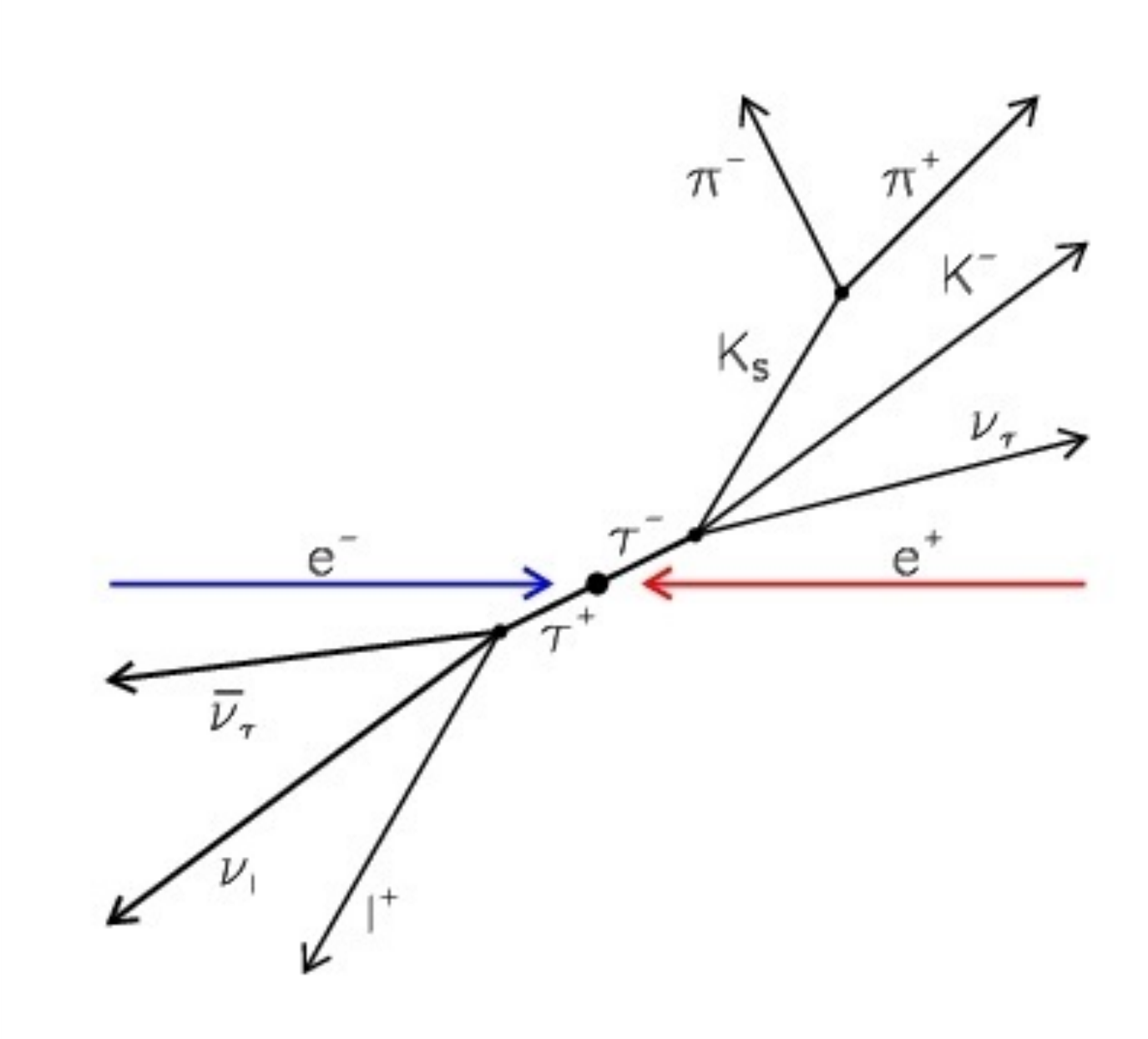}\\
\parbox[h]{0.45\textwidth}{\caption 
{Schematic view of the $\tau$ decay chains in
$e^+e^-\to\tau^+\tau^-$ events selected for this
analysis. Lepton $l^+$ can be electron or muon.  } 
\label{decays}} \hfill
\end{figure}

The branching fraction for the $\tau^{-}\to K^{-}K_S\nu_{\tau}$ decay has been
measured with relatively high (3\%) precision by the Belle 
experiment~\cite{Belle}. The $K^-K_S$ mass spectrum 
was measured by the CLEO experiment~\cite{CLEOt}. In the CLEO analysis, 
a data set of $2.7\times10^6$ produced $\tau$ pairs was used, and 
about 100  events in the decay channel $\tau^-\to K^-K_S\nu_{\tau}$
were selected.
In this work, using  about $\sim$ 10$^9$   $\tau$ leptons, 
we significantly improve upon the measurement of the spectral function for 
the $\tau^-\to K^-K_S\nu_{\tau}$ decay.

\section{Data used in the analysis}
We analyze a data sample corresponding to an integrated
luminosity of 468 fb$^{-1}$ recorded with the \babar\
detector~\cite{Babar1},\cite{Babar2} at the SLAC PEP-II asymmetric-energy $e^+e^-$ collider.

For simulation of $e^+e^-\to \tau^+\tau^-$ events 
the KK2f Monte Carlo generator~\cite{kk2f} is used, which includes
higher-order radiative corrections to the Born-level process.
Decays of $\tau$ leptons are simulated using  the 
Tauola package~\cite{Tauola}.
Two separate samples of simulated $e^+e^-\to \tau^+\tau^-$ events are used: 
a generic sample with $\tau$ decaying to all significant final states, 
and the signal channel where 
$\tau^+\to l^+\nu_l\bar{\nu}_\tau$, $l=e$ or $\mu$ and 
$\tau^-\to K^-K_S\nu_\tau$.
To estimate backgrounds, we use a sample of simulated generic  
$e^+e^-\to \tau^+\tau^-$ events after excluding the signal decay channel 
($\tau^+\tau^-$ background) and a sample containing all events arising from 
$e^+e^-\to q\bar{q}$, $q=u,d,s,c$ and $e^+e^-\to B\bar{B}$ processes 
($q\bar{q}$ background).
The $q\bar{q}$ background  events with $q=u,d,s,c$
are generated using the JETSET generator~\cite{JETSET}, while $B\bar{B}$ 
events are simulated with EVTGEN~\cite{EVTGEN}. The detector response is 
simulated with GEANT4 ~\cite{GEANT4}. The equivalent luminosity of the 
simulated sample is 2-3 times higher than the  integrated luminosity in data.

\section{Event selection\label{Sel}}
\begin{figure}
\includegraphics[width=0.45\textwidth]{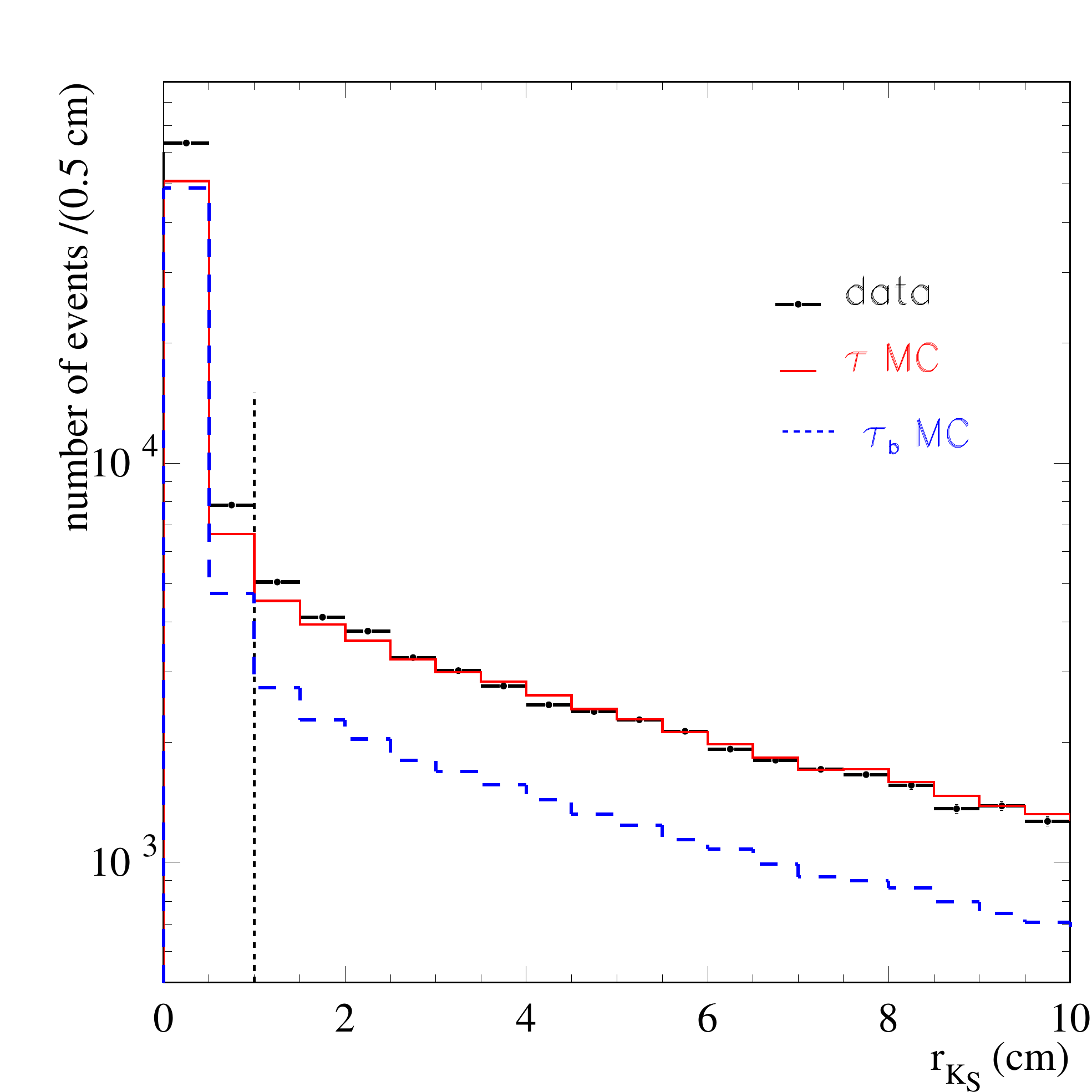} \\
\parbox[h]{0.45\textwidth}{\caption
{The $K_S$ candidates decay length distribution for data (points with errors),
$\tau^+\tau^-$ simulation events (solid histogram), 
and $\tau$ background simulation (dashed
histogram). 
The vertical line indicates the boundary of the selection condition.}
\label{rkscut}} \hfill
\end{figure}

\begin{figure}
\includegraphics[width=0.45\textwidth]{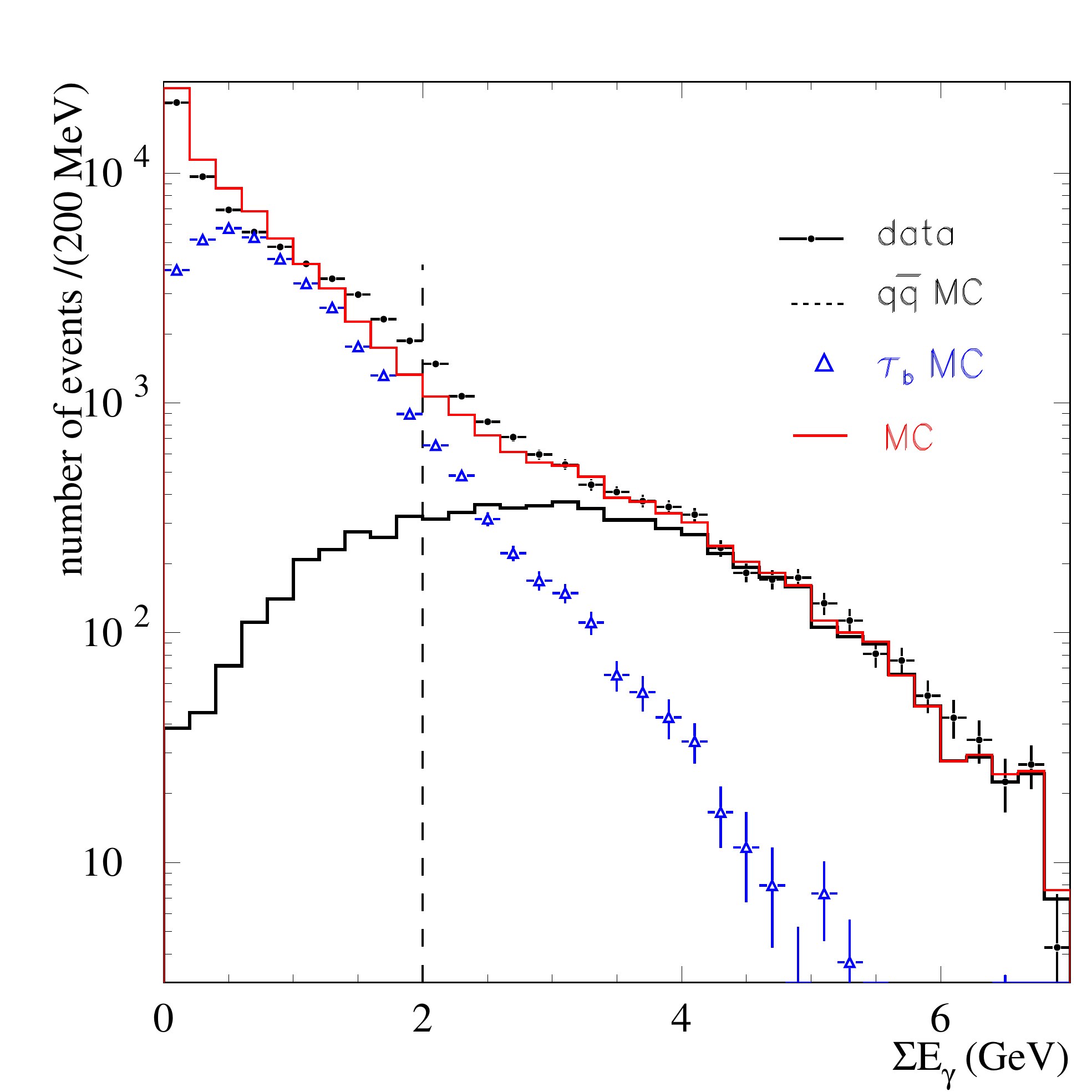} \\
\parbox[h]{0.45\textwidth}{\caption
{Distributions of the total energy of photons in the event
for data (points with errors), 
$\tau^+\tau^-$  and $q\bar{q}$  simulation events (solid histogram), 
$\tau$ background 
simulation (empty triangles with errors) and $q\bar{q}$
background simulation (dashed histogram). 
The vertical line indicates the boundary 
of the selection condition.}
\label{sumeg1}} \hfill
\end{figure}

We select $e^+e^-\to \tau^+\tau^-$ events with  the $\tau^+$  decaying
leptonically ($\tau^+\to l^+\nu_l\bar{\nu}_\tau$, $l=e$ or $\mu$) and the
$\tau^-$ decaying to $K^-K_S\nu_\tau$.
Such events referred to as signal events below.
The $K_S$ candidate is detected in the
$K_S\to\pi^+\pi^-$ decay mode. The topology of events to be selected is
shown in Fig.~\ref{decays}. Unless otherwise stated, all
quantities are measured in the laboratory frame.
The selected events must satisfy the following requirements:
\begin{itemize}
\renewcommand{\labelitemi}{--}
\item The total number of charged tracks, $N_{\rm trk}$,  must be four 
and the total charge of the event must be zero.

\item  Among the four charged tracks there must be an identified lepton   
(electron or muon) and an identified kaon of opposite charge. 
The track origin point 
requirements are $|d_0| <$ 1.5 cm and $|z_0| <$ 2.5 cm, 
where $|d_0|$ and $|z_0|$ are the
distances between the track and the interaction region center in transverse 
and longitudinal directions with respect to the beams.   

\item To reject $\mu$ pairs and Bhabha events,
the lepton candidate must have a momentum above 1.2 GeV/$c$, the momentum in
the center-of-mass frame (c.m. momentum) must be smaller than 4.5 GeV/$c$,
and the cosine of the lepton polar angle
$|\cos{\theta_{l}}|$  must be below 0.9. 

\item  To suppress background from charged pions,
the charged kaon candidate must have a momentum, $p_{K}$,  above 0.4 GeV/$c$
and below 5 GeV/$c$, and the cosine of its polar angle must lie between -0.7374
and 0.9005.

\item The two remaining tracks, assumed to be pions,
form the $K_S$ candidate. The $\pi^+\pi^-$ invariant mass must lie 
within 25 MeV/$c^2$ of the nominal $K_S$ mass, 497.6 MeV/$c^2$. The 
$K_S$ flight length $r_{K_S}$, measured as the distance between
the $\pi^+\pi^-$ vertex and the collision point, must be larger than 1 cm.
The $r_{K_S}$ distributions for data events and simulated signal events
are shown in Fig.~\ref{rkscut}.

\item  The total energy in neutral clusters, $\Sigma E_{\gamma}$, must be less
than 2 GeV (Fig.~\ref{sumeg1}). 
Here, a neutral cluster is defined as a local energy deposit
in the calorimeter with energy above 20 MeV and no associated charged
track.

\item The magnitude of the thrust \cite{thrust1,thrust2}
for the event,  calculated using 
charged tracks only,   must be greater than 0.875.

\item  The angle defined by the momentum of the lepton and that 
of the $K^-K_S$ system in the c.m.\ frame must be larger than
110 degrees.

\end{itemize} As a result of applying these selection criteria
the $\tau$ background is suppressed by 3.5 orders of magnitude,
and the $q\bar{q}$ background by 5.5 orders.

\begin{figure}
\includegraphics[width=0.45\textwidth]{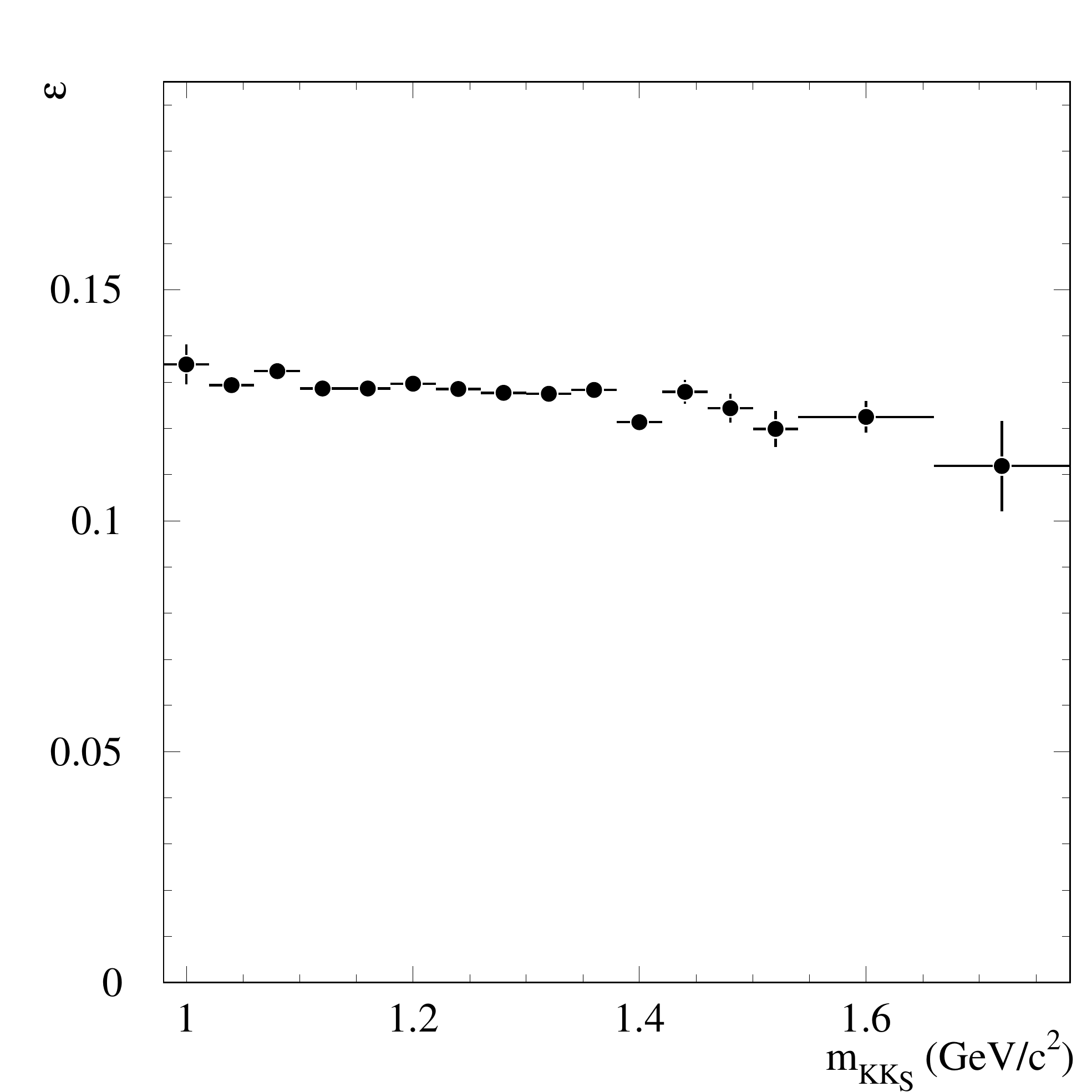} \hfill
\parbox[h]{0.45\textwidth}{\caption { 
Selection efficiency as a function of the $K^-K_S$ invariant
mass, according to simulation. }
\label{seleff1}} \hfill
\end{figure}
\begin{figure}
\includegraphics[width=0.45\textwidth]{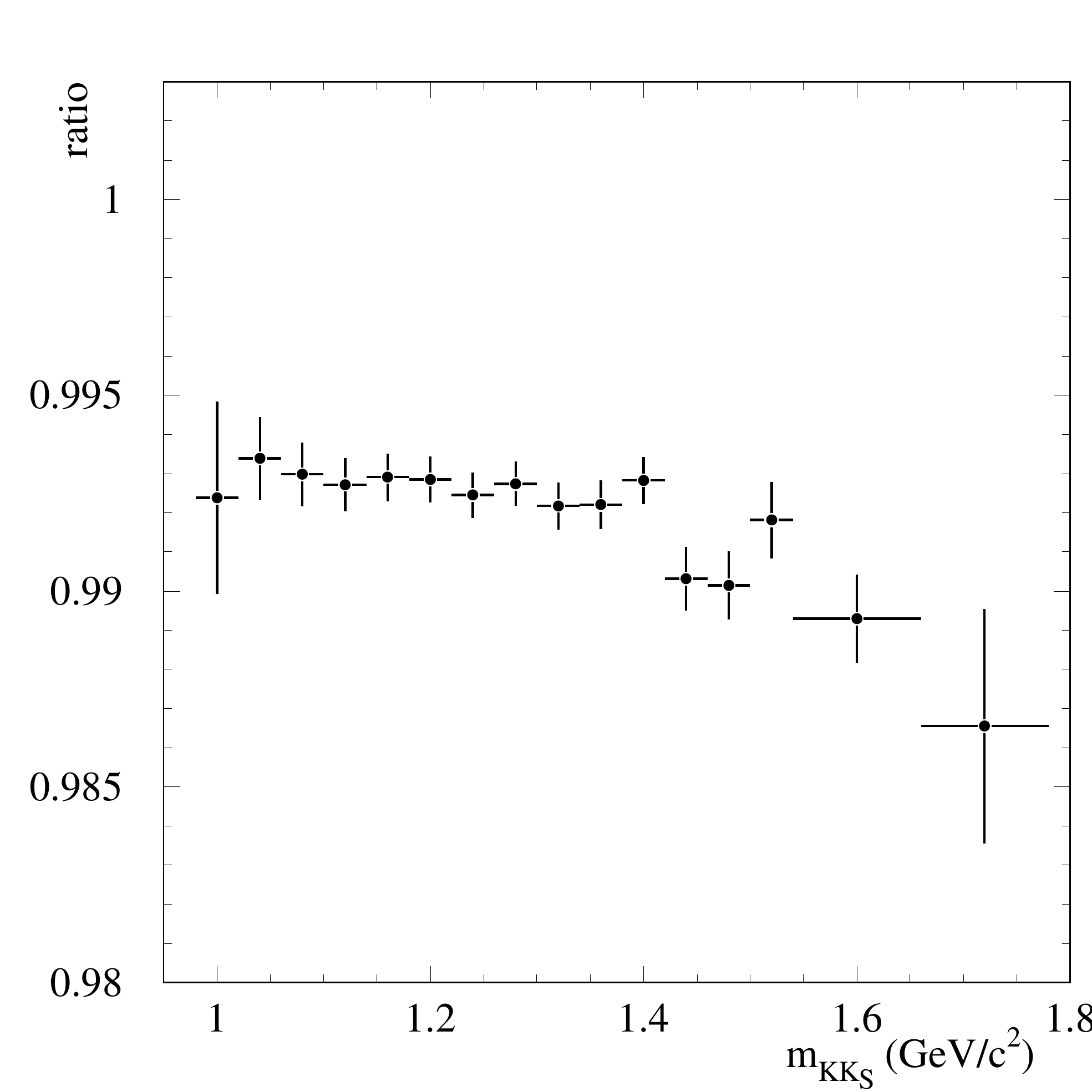}
\parbox[h]{0.45\textwidth}{\caption {Efficiency correction 
factor for adjusting the simulation PID
efficiency to match the efficiency measured on data,
as a function of the $K^-K_S$ mass for signal events. }
\label{pcorr}} 
\end{figure}

\begin{figure}
\includegraphics[width=0.48\textwidth]{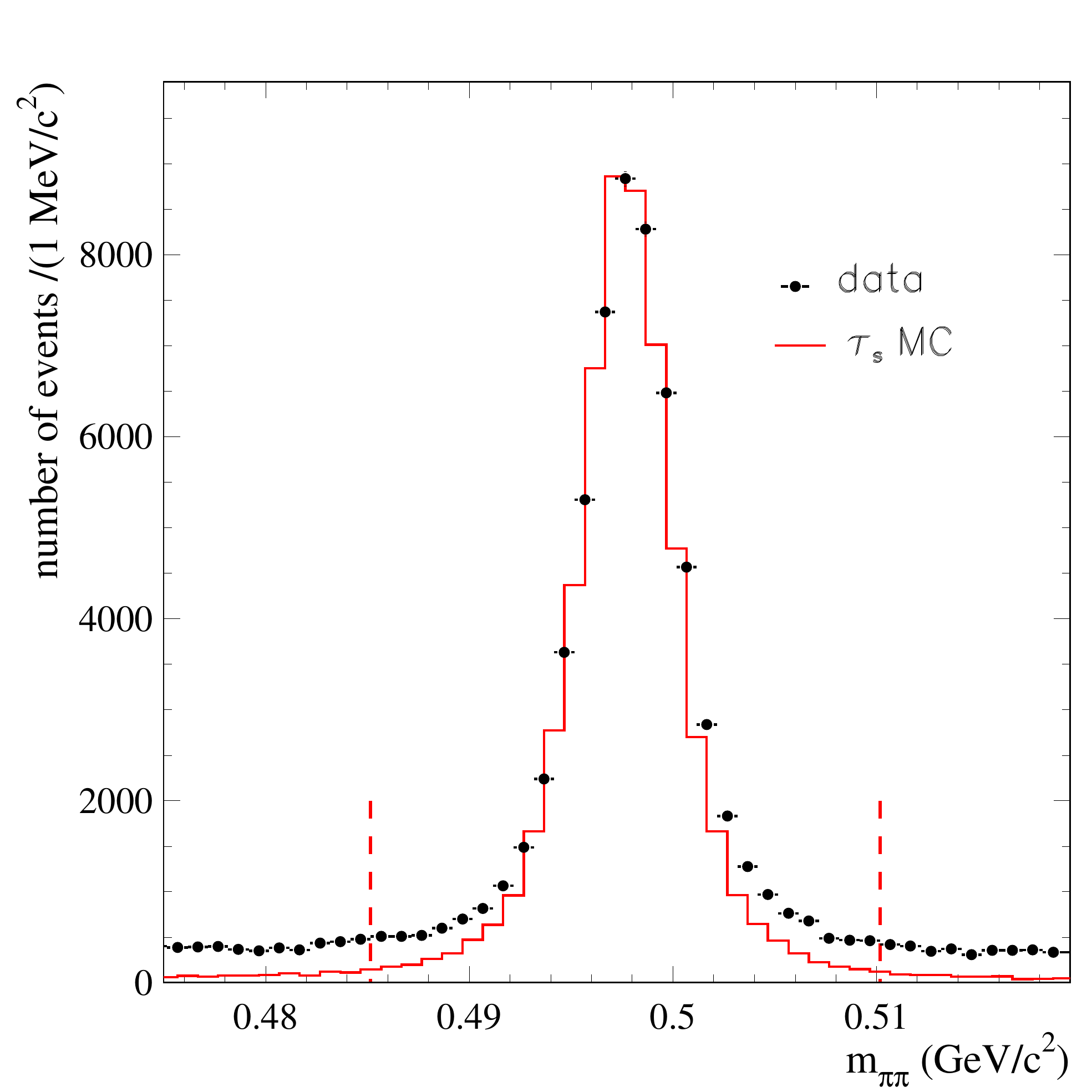} \hfill
\caption { The $\pi^+\pi^-$ mass spectrum for $K_S$
candidates in data (points with errors) and signal
simulation (histogram).
Between the two vertical lines there is a signal region
used in the procedure of non-$K_S$ background subtraction.
\label{ksmas}}
\end{figure}

\section{Detection efficiency\label{deteff}}
The detection efficiency obtained after applying the selection criteria is
calculated using signal Monte Carlo simulation as a function of the 
true $m_{\rm K^-K_S}$
mass and is shown in Fig.~\ref{seleff1}. 
The efficiency is 
weakly dependent on $m_{K^-K_S}$. The average 
efficiency over the mass spectrum is about 13\%. 
It should be noted that the  $K^-K_S$ mass resolution is 
about 2-3 MeV/$c^2$,  significantly smaller than the size of the mass bin
(40 MeV/$c^2$) used in Fig.~\ref{seleff1}.  
Therefore, in the following we neglect the effects of the finite 
$K^-K_S$   mass resolution.

To correct for the imperfect simulation of the kaon identification
requirement, 
the particle identification (PID) efficiences have been compared for
data and simulation on high purity control samples of kaons from 
$D^{\star}\to\pi^+D^0,~D^0\to K^-\pi^+$ decays \cite{BBPhys}.
We correct
the simulated efficiency using the measured ratios of the
efficiencies measured in data and Monte Carlo, in bins of the kaon
candidate momentum and polar angle. The resulting correction factor as
a function of $m_{K^-K_S}$ is shown in Fig.~\ref{pcorr}.

\section{Subtraction of non-$\bf{K_S}$ background\label{specks}}
The  $\pi^+\pi^-$ mass spectra for $K_S$ candidates in data 
and simulated signal events are shown in Fig.~\ref{ksmas}.
The data spectrum consists of a peak at the $K_S$ mass and a flat 
background. 
To subtract the non-$K_S$ background, the following procedure is used.
The signal region is set to $\pi^+\pi^-$ masses within 0.0125 GeV/$c^2$ of the
$K_S$ mass (indicated by arrows in  Fig.~\ref{ksmas}), 
and the sidebands are set to between 
0.0125 and 0.0250 GeV/$c^2$ away from the nominal
$K_S$ mass. Let $\beta$ be the
fraction of events with a true $K_S$ that fall in the
sidebands, and let $\alpha$ be the fraction of  non-$K_S$ events
that fall in the sidebands. The total number of events in the signal
region plus the sidebands, $N$, and the number of events in the
sidebands, $N_{sb}$, depend on  the number 
of true $K_S$, $N_{K_S}$,
and the number of non-$K_S$ background events, $N_{b}$
according to the following relation :

\begin{subequations}
\begin{equation} N = N_{K_S} + N_{b},
\end{equation}
\begin{gather} N_{sb} = \alpha\cdot N_{b} + \beta\cdot N_{K_S}
\end{gather}
\end{subequations}

Therefore:
\begin{equation}
N_{K_S}=(\alpha N - N_{sb})/(\alpha-\beta).
\label{eq6}
\end{equation}
The value of $\beta$ is determined  using $\tau$  signal simulation. 
It is found to be nearly  independent of the $m_{K^-K_S}$ 
mass and is equal to   0.0315~$\pm$~0.0015.
The value of  $\alpha$ is expected to be 0.5 for a uniformly distributed 
background.  This  is   
consistent with the value 0.499~$\pm$~0.005 obtained on simulated $\tau^+\tau^-$
background events. 
The non-$K_S$ background is subtracted in each $m_{K^-K_S}$ bin. Its
fraction is found to be about 10\% of the selected events with  
$m_{K^-K_S}$ near and below 1.3 GeV/$c^2$ and increases up to 50\% above
1.6 GeV/$c^2$. 
\begin{figure}
\includegraphics[width=0.45\textwidth]{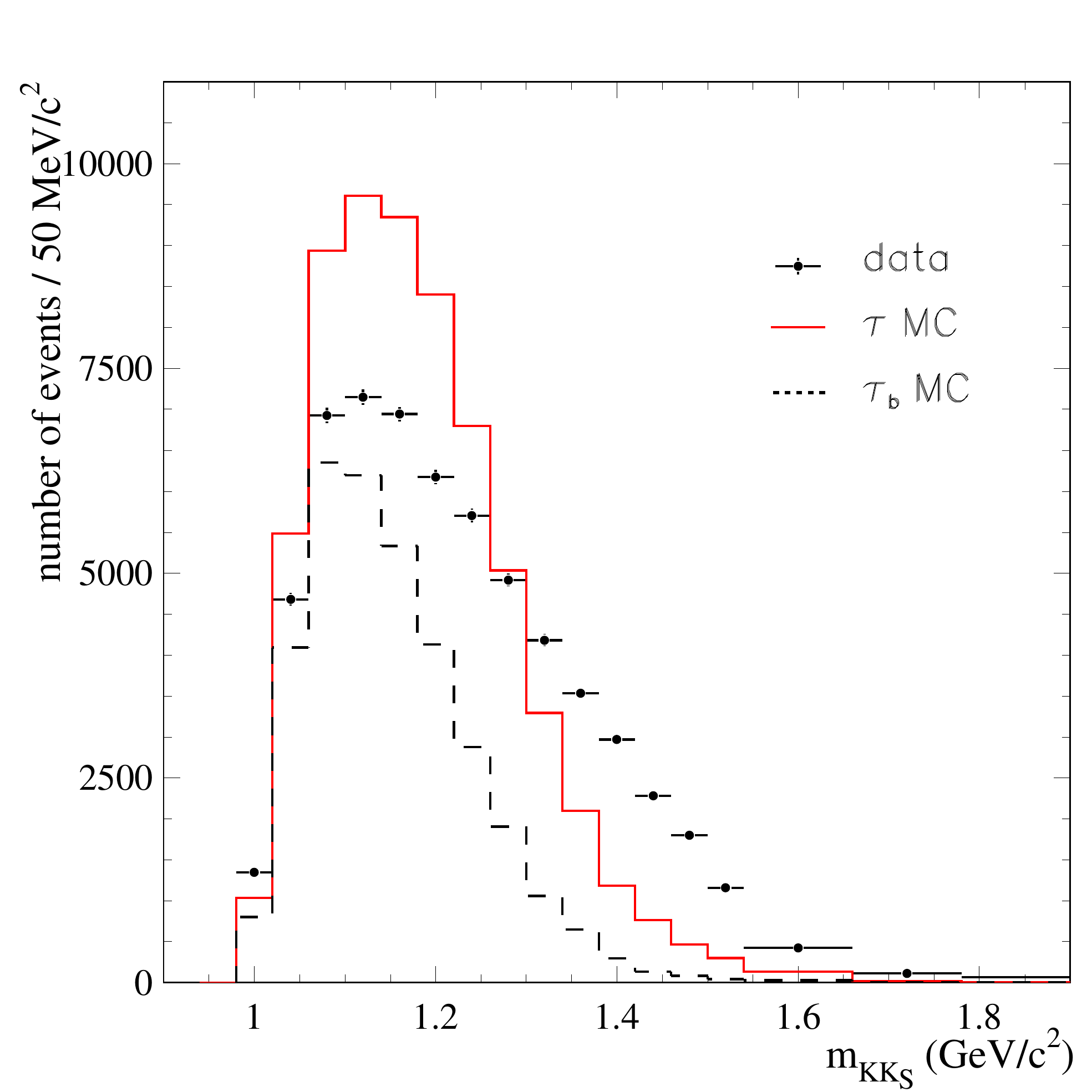}
\parbox[h]{0.45\textwidth}{\caption { The 
$m_{K^-K_S}$ spectra for data (points with errors),
$\tau^+\tau^-$ simulation events (solid histogram),
and $\tau$ background simulation (dashed histogram).
The non-$K_S$ background is subtracted. }
\label{kksm}} \hfill
\end{figure}

\section{
Subtraction of $\tau$-background with a $\pi^0$\label{rmspe}}
Although the studied process $\tau^-\to K^-K_S\nu_{\tau}$
is not supposed to contain a $\pi^0$ in the final state,   
some events from background processes with a $\pi^0$ pass the selection 
criteria. In the following, we describe how the $\pi^0$ background 
contribution is subtracted.

The  $K^-K_S$ mass spectra for selected data and $\tau^+\tau^-$
simulated events after
subtraction of the non-$K_S$  background are shown in Fig.~\ref{kksm}.
According to the simulation, the number of signal and
$\tau$-background events are of the same order of magnitude. 
The $\tau^+\tau^-$ background 
consists of events with the decay $\tau^-\to K^-K_S\pi^0\nu_{\tau}$ 
(79\%), events with a misidentified kaon from decays    
$\tau^-\to \pi^-K_S\nu_{\tau}$ (10\%) and 
$\tau^-\to \pi^-K_S\pi^0 \nu_{\tau}$ (3\%), 
and events with a misidentified lepton mainly from the decays
$\tau^+\to \pi^+\bar{\nu}_{\tau}$ and
$\tau^+\to \pi^+\pi^0\bar{\nu}_{\tau}$ (7\%).
Thus, more than 80\% of the background events contain a $\pi^0$ in the final state. 
It should be noted  that events with a misidentified lepton
have the same $m_{K^-K_S}$ distribution as signal events.

The branching fractions for the background modes without a $\pi^0$,
$\tau^-\to \pi^-K_S\nu_{\tau}$ and $\tau^+\to \pi^+\bar{\nu}_{\tau}$, 
have been measured with high precision (1.7\% and 0.5\%)~\cite{PDG}. The
hadronic mass spectrum for $\tau^-\to \pi^-K_S\nu_{\tau}$ is also well
known~\cite{piks} and this decay  proceeds mainly 
via the $K^\ast(892)$ intermediate state.
Therefore all $\tau^+\tau^-$ background without a $\pi^0$
is subtracted using Monte Carlo simulation. The amount of $q\bar{q}$ background,
not shown in Fig.~\ref{kksm},  is about 2\% of selected data events.
The part of this background without a $\pi^0$ is also
 subtracted using Monte Carlo simulation.

The branching fractions for the background modes 
$\tau^-\to K^-K_S\pi^0\nu_{\tau}$,
$\tau^-\to \pi^-K_S\pi^0 \nu_{\tau}$, and
$\tau^+\to \pi^+\pi^0\bar{\nu}_{\tau}$ are measured with
a precision of 4.7\%, 3.4\%, and 0.4\%, respectively. The hadronic mass spectrum
is well known only for  the last decay~\cite{pipi0}.
For the two other decays, 
only low-statistics measurements~\cite{CLEOt} are available. 
Therefore, we use the data to subtract the $\tau$-background with $\pi^0$
from the  $K^-K_S$ mass spectrum.  To do this, the selected events are
divided into two classes, without and with a  
$\pi^0$ candidate, which is
defined as a pair of photons with an invariant mass in the range 
$100-160$ MeV/$c^2$. 

\begin{figure}
\centering
\includegraphics[width=0.45\textwidth]{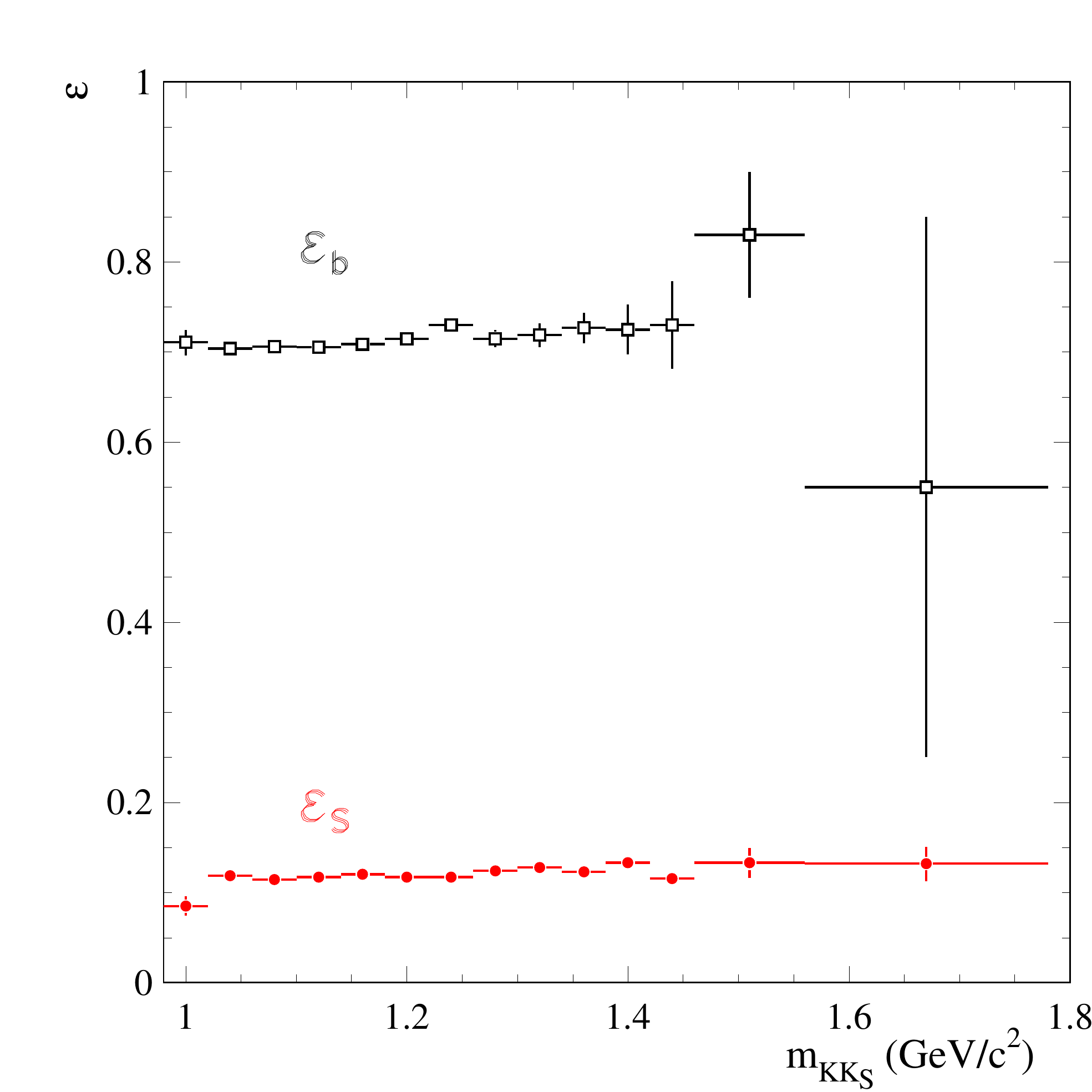} 
\caption {The probabilities $\epsilon_{ s}$ and
$\epsilon_{ b}$ used in Eqs.~(\ref{eq8}a,\ref{eq8}b) as functions of the 
$K^-K_S$ mass,measured on simulated events.
\label{efcut}}
\end{figure}

On the resulting sample, the numbers of signal ($N_{s}$) and background
$\tau^+\tau^-$ events containing a $\pi^0$ candidate ($N_{ b}$) are obtained
in each $m_{ K^-K_S}$ bin:
\begin{subequations}
\begin{equation}
N_{0\pi^0}=(1-\epsilon_{s})N_{s}+(1-\epsilon_{b})N_{b},
\end{equation}
\begin{gather}
N_{1\pi^0}=\epsilon_{s}N_{s}+\epsilon_{b}N_{b},
\end{gather}
\label{eq8}
\end{subequations}
where $N_{0\pi^0}$ and $N_{1\pi^0}$ are the numbers of selected data events
with zero and at least one $\pi^0$ candidate, and 
$\epsilon_{ s}$ ($\epsilon_{ b}$) is the probability for signal  
(background) $\tau^+\tau^-$ events to be found in events 
with at least one $\pi^0$ candidate calculated using Monte Carlo simulation. 
The values  $\epsilon_{ s}$ and $\epsilon_{ b}$ for each bin 
in  $m_{ K^-K_S}$ are measured in Monte
Carlo by counting how many signal and background event candidates
contain a $\pi^0$ candidate.
Fig.~\ref{efcut} shows the $\epsilon_{ s}$ and $\epsilon_{ b}$ measured 
in Monte Carlo as a function of $m_{ K^-K_S}$.
The efficiency $\epsilon_{ b}$ is corrected to take into account 
the different $\pi^0$ efficiency between data and Monte Carlo
as measured on data and simulated control samples in the ISR 
$e^+e^-\to \omega(783)\gamma \to \pi^+\pi^-\pi^0\gamma $  process \cite{pi0ef}. 
The average correction is  $\delta=0.976\pm0.008$.
\begin{figure}
\centering
\includegraphics[width=0.45\textwidth]{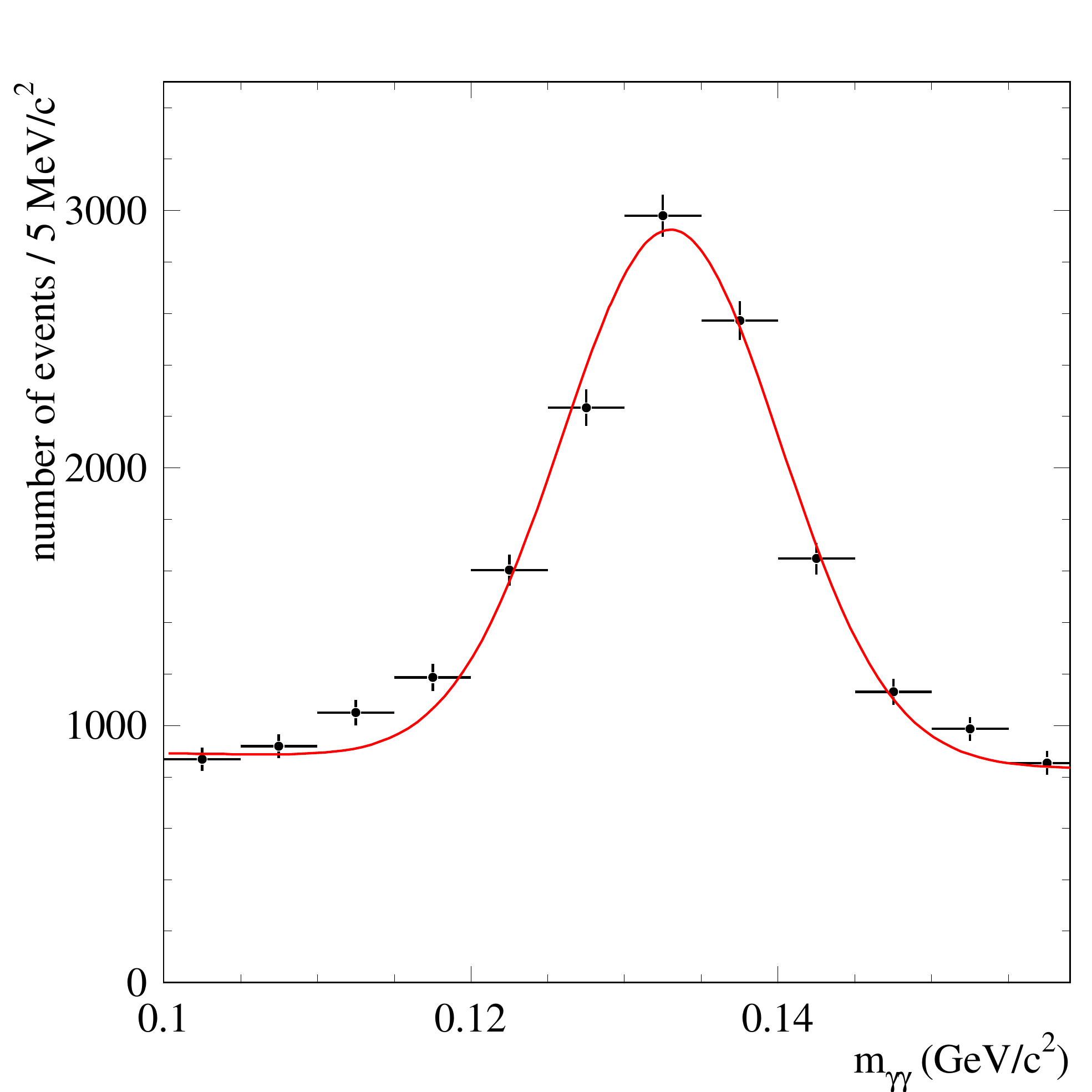} 
\caption {Two-photon invariant mass spectrum  of $\pi^0$ candidates
in data. The curve corresponds to the fit function,
described in text. \label{pi0bg}}
\end{figure}
The non-zero value of  $\epsilon_{ s}$ is due to random
combinations of two spurious photons originating from beam background
or nuclear interactions of charged kaons or pions.
The beam-generated background is simulated
by using special background events recorded 
during normal data-taking conditions but with a 
randomly generated  trigger. These events are superimposed on simulated
events. 
The following procedure is used to measure $\epsilon_{ s}$ on data events.
We compare the 
solution of Eqs.~(\ref{eq8}a,\ref{eq8}b) described 
above with the solution of the same 
system, in which the number of events with $\pi^0$ is determined from the 
fit to the two-photon invariant mass spectrum of $\pi^0$ candidates.
Since the mass dependence of $\epsilon_{ s}$ and $\epsilon_{ b}$ 
is mild (Fig.~\ref{efcut}), this
comparison is performed using the full sample of selected events 
without splitting the sample
into $K^-K_S$ mass bins. 
The two-photon mass spectrum of $\pi^0$ candidates in data is shown 
in Fig.~\ref{pi0bg}.
\begin{figure}
\centering
\includegraphics[width=0.45\textwidth]{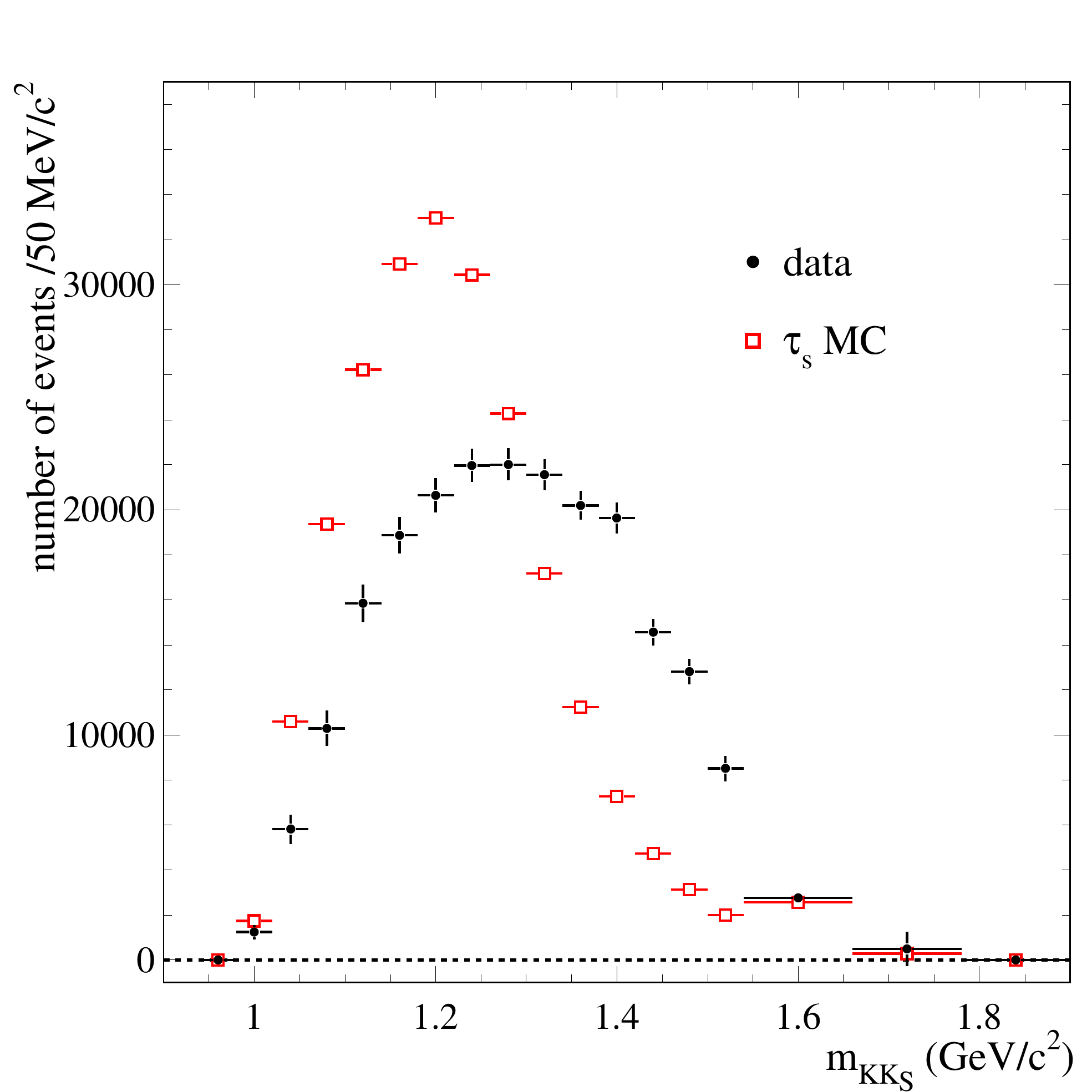} \hfill
\includegraphics[width=0.45\textwidth]{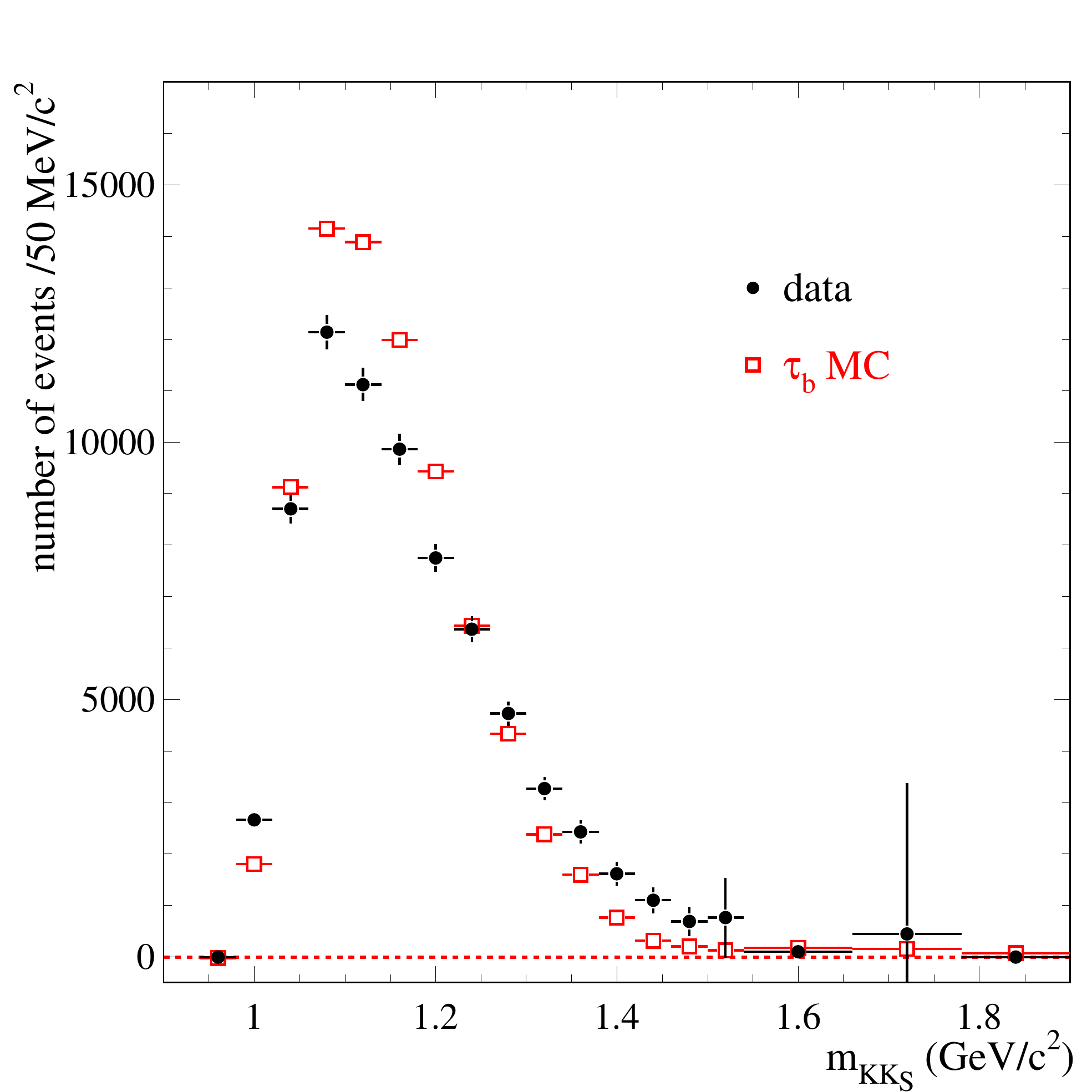} 
\caption { Measured $m_{ K^-K_S}$ spectra for signal events (top) 
and $\tau$ background events with $\pi^0$ (bottom) in comparison with 
the $\tau$ signal ($\tau_S$) and  $\tau$ background ($\tau_b$)   
Monte Carlo  simulation.
\label{dtmc}}
\end{figure}

The spectrum in Fig.~\ref{pi0bg} is fitted by a sum 
of a Gaussian and a flat component.
The numbers $N_{1\pi^0}$ and $N_{0\pi^0}$ on the left side of 
Eqs.~(\ref{eq8}a,\ref{eq8}b)
are substituted by $N^{\star}_{1\pi^0}=N_{1\pi^0}-N^{\rm lin}_{1\pi^0}$ and
$N^{\star}_{0\pi^0}=N_{0\pi^0}+N^{\rm lin}_{1\pi^0}$, where $N^{\rm lin}_{1\pi^0}$
is the number of events under the  flat component,
 obtained after fitting the $\gamma\gamma$  spectrum in Fig.~\ref{pi0bg}. 
 The  value $\epsilon_b$ is substituted by
$\epsilon^{\star}_b=w\epsilon_b$, where $w=0.682\pm 0.010$  
is the fraction of events with a reconstructed  $\pi^0$ for simulated
$\tau^+\tau^-$ background  
(Fig.~\ref{pi0bg}). The term `reconstructed $\pi^0$'  corresponds 
to $\pi^0$s in the Gaussian part in  Fig.~\ref{pi0bg}.
The modified system of equations is:
\begin{subequations}
\begin{equation}
N^{\star}_{0\pi^0}=N_{s}+(1-\epsilon^{\star}_{b})N_{b},
\end{equation}
\begin{gather}
N^{\star}_{1\pi^0}=\epsilon^{\star}_{b}N_{b}.
\end{gather}
\label{eq17}
\end{subequations}
In  Eqs.~(\ref{eq17}a,\ref{eq17}b) 
the top line contains all events without a reconstructed
$\pi^0$, while the lower line contains events with at least one 
reconstructed  $\pi^0$.
After subtracting the spurious $\pi^0$s corresponding to the flat
background in Fig.~\ref{pi0bg}, Eqs.~(\ref{eq17}a,\ref{eq17}b) no longer 
contains $\epsilon_s$ nor a contribution from the $\pi^0$
background.

The average value of $\epsilon_{b}$ from  Fig.~\ref{efcut}
is $0.720\pm 0.003$,  giving
$\epsilon^{\star}_{b}=0.491\pm 0.008$ on the average.  
This  value is then corrected by the reconstructed
$\pi^0$ efficiency correction factor $\delta$,
as discussed above. 
The number of signal events, 
$N_s$,  obtained by solving  Eq.~(\ref{eq17}a,\ref{eq17}b)  
and using the corrected value of $\epsilon^{\star}_{ b}$
is about 1\% higher than the
previous one, derived from Eq.~(\ref{eq8}a,\ref{eq8}b). 
This 1\% shift in $N_{ s}$ is explained by 
the difference between data and Monte Carlo simulation in $\epsilon_{ s}$.

  To obtain the final $K^-K_S$ mass spectrum  
we return to  Eqs.~(\ref{eq8}a,\ref{eq8}b). 
Based on the above study of the $\pi^0$ systematics we must 
correct the efficiencies $\epsilon_{ s}$ and $\epsilon_{ b}$. 
First, we correct the value of $\epsilon_{ b}$ by the $\pi^0$ 
efficiency correction $1-w(1-\delta)\simeq 0.984\pm 0.006$, 
where  $w$ and $\delta$  are defined above.
Then we adjust the value of $\epsilon_{ s}$ by a factor $1.05\pm0.05$
to take into account the above-mentioned 1\% correction in flat
background simulation. 
Then the number of simulated $\tau^+\tau^-$ background events without 
a $\pi^0$ is multiplied by a factor of
 $p=0.92\pm0.02$ to take into account the difference between experimental
$\tau$  branching fractions and branching fractions used in the Tauola Monte Carlo
generator.  With  these corrected values for 
$\epsilon_{ s}$ and $\epsilon_{ b}$ we solve Eqs.~(\ref{eq8}a,\ref{eq8}b) for each 
$K^-K_S$ mass bin and obtain mass spectra for signal ($N_{ s}$) 
and background ($N_{ b}$).

 The efficiency corrected signal mass spectrum, using the signal efficiency
from Fig.~\ref{seleff1}, is shown in Fig.~\ref{dtmc}(top), 
in comparison with the simulation.
The $\tau$-pair   $m_{ K^-K_ S}$  background spectrum 
(Fig.~\ref{dtmc}(bottom)) is 
compared with simulation without efficiency correction.
Spectra are normalized to the same number of events.
We find a substantial difference between data and simulation for
the signal spectrum, and better agreement for the background spectrum.

\section{ Systematic uncertainties }
\begin{table}
\caption{The systematic uncertainties on 
{\cal B}($\tau^-\to K^-K_S\nu_{\tau}$) from different sources.
\label{tab-sys}}
\begin{ruledtabular}
\begin{tabular}{lc}
Sources & uncertainty (\%)\\ \hline
Luminosity  & 0.5 \\
Tracking efficiency & 1.0 \\
PID   & 0.5\\
non-$K_S$ background subtraction & 0.4\\
$\tau^+\tau^-$ background without $\pi^0$ & 0.3\\
$\tau^+\tau^-$ background with $\pi^0$ & 2.3\\
$q\bar{q}$ background & 0.5\\ \hline
total & 2.7\\
\end{tabular}
\end{ruledtabular}
\end{table}

This section lists all the uncertainties  in the parameters 
used in this analysis, and estimates the  overall systematic 
uncertainty on the $\tau^-\to K^-K_S\nu_{\tau}$ branching
fraction and the $K^-K_S$ mass spectrum.

\begin{figure}
\includegraphics[width=0.45\textwidth]{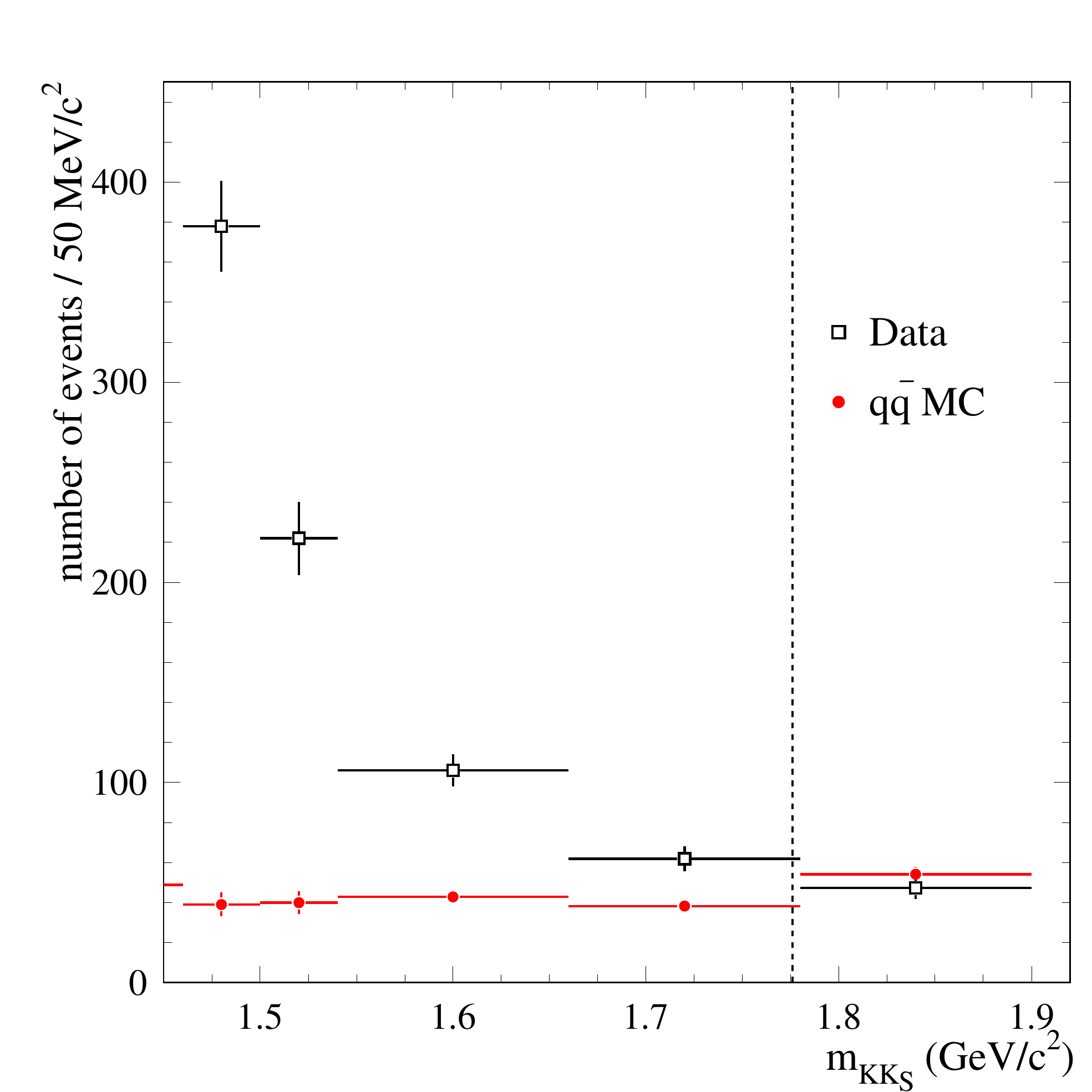} \hfill
\includegraphics[width=0.45\textwidth]{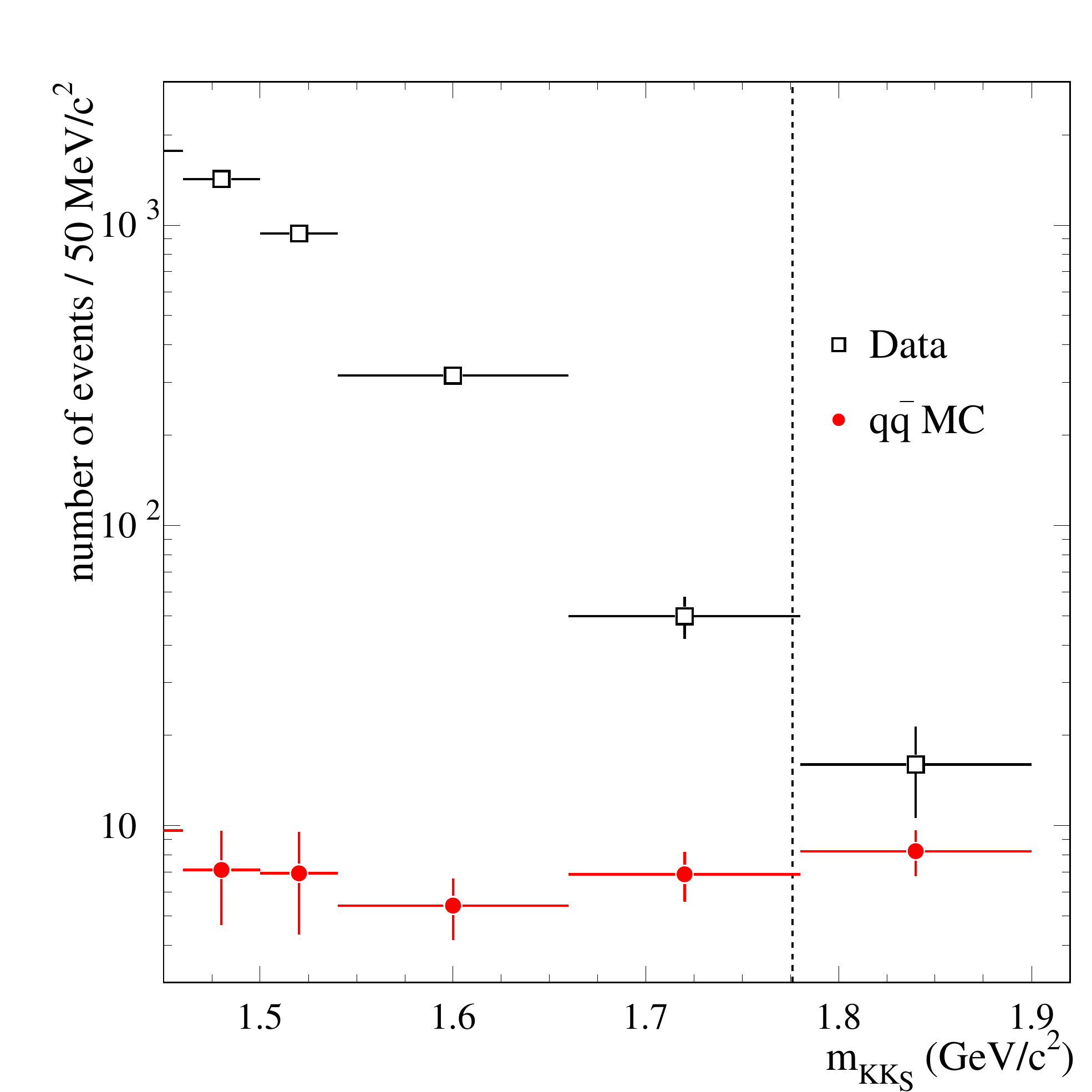} 
\caption{
$K^-K_S$ mass spectra near the end point $M_{ K^-K_S}=m_\tau$
for selected data and $q\bar{q}$ simulated events 
without (top) and with (bottom) a  $\pi^0$ candidate. 
 The vertical line indicates the $\tau$ mass.
\label{qbrg}}
\end{figure}
\begin{figure}
\centering
\includegraphics[width=0.45\textwidth]{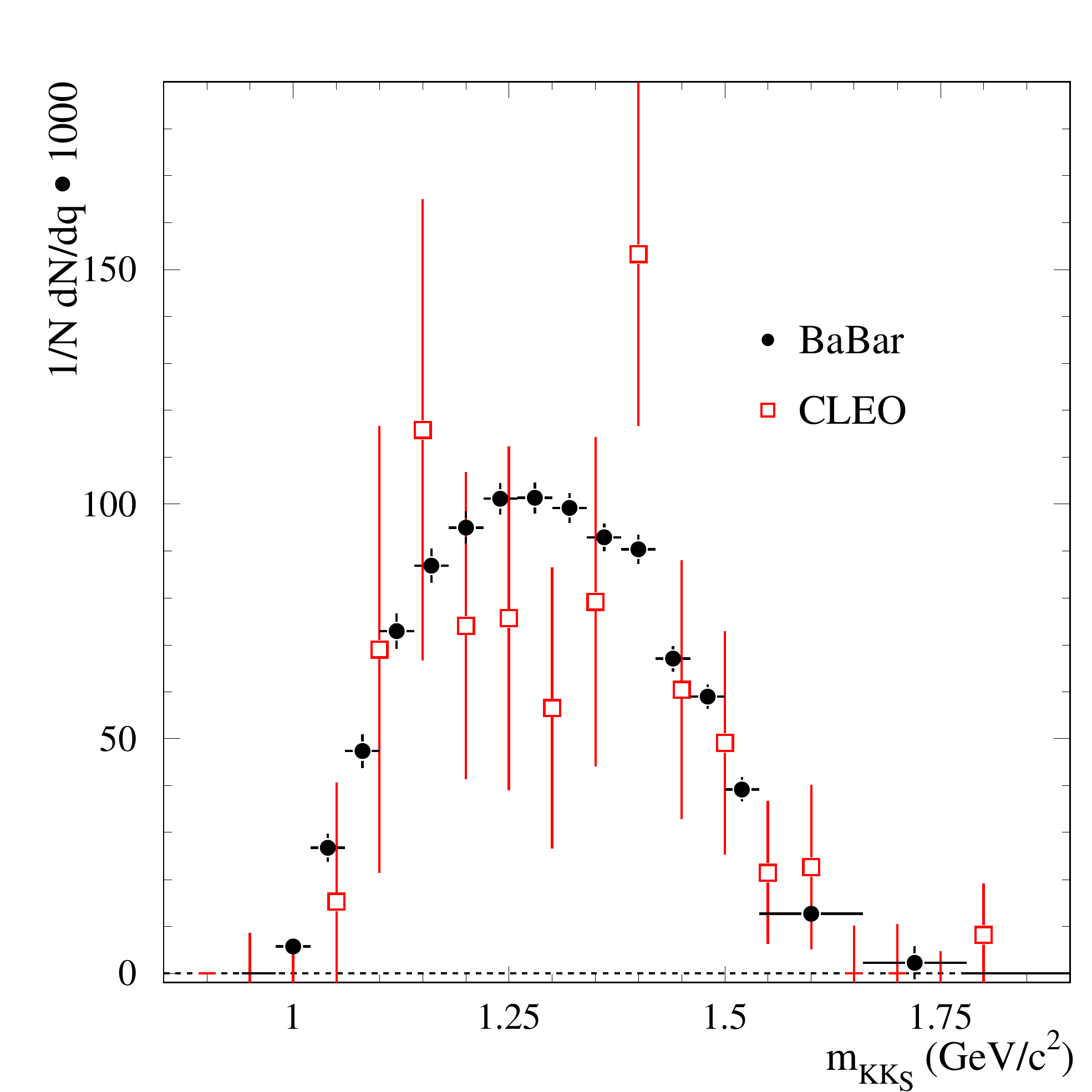}
\caption { Normalized $K^-K_S$ invariant mass
spectrum for the $\tau^-\to K^-K_S\nu_{\tau}$ decay measured in this work
(filled circles) compared to the CLEO measurement~\cite{CLEOt} (empty squares).
Only statistical uncertainties are shown.
\label{mspec}}
\end{figure}
  The subtraction of non-$K_S$ background  is described in Section~\ref{specks}.
To check the procedure of the non-$K_S$ background subtraction, 
we varied the coefficients of $\alpha$ and $\beta$
within their uncertainties, which leads to a systematic 
uncertainty  of 0.4\% in
the $\tau^-\to K^-K_S\nu_{\tau}$ branching fraction.
This uncertainty is independent of the $K^-K_S$ mass.

The PID corrections were discussed in Section~\ref{deteff}. The systematic
uncertainty due to data-Monte Carlo simulation difference
in particle identification
is taken to be 0.5\%, independent of the $K^-K_S$ mass.
The uncertainty on how well the Monte Carlo simulates the tracking
efficiency is estimated to be 1\%.

Fig.~\ref{qbrg} shows the $m_{ K^-K_S}$ spectra for selected data events 
with and without a $\pi^0$ candidate  near the 
end point $m_{ K^-K_S}=m_\tau$ compared to simulated $q\bar{q}$ events. 
It appears that the number of data and simulated $q\bar{q}$ events are in 
reasonable agreement at $m_{ K^-K_S} > m_{\tau}$, where all data events are 
expected to be from the $q\bar{q}$ background.
We take the observed difference between data and Monte Carlo 
near the end point $M_{ K^-K_S}=m_\tau$ in Fig.~\ref{qbrg}
as an uncertainty on the $q\bar{q}$ background. This leads to an uncertainty 
on ${\cal B}(\tau^-\to K^-K_S\nu_{\tau})$ of 0.5\%. 

The uncertainty associated with the subtraction of the $\tau^+\tau^-$ background
with  $\pi^0$'s is estimated by varying the efficiencies
$\epsilon_s$ 
and $\epsilon_b$ used in Eqs.~(\ref{eq8}a,\ref{eq8}b) within their systematic 
uncertainties: 5\% in $\epsilon_s$ 
(uncertainty in the number of  spurious $\pi^0$s) 
and 6\% in $\epsilon_b$ (uncertainty in numbers of both 
spurious and reconstructed $\pi^0$s).
The corresponding contribution to the systematic uncertainty on
${\cal B}(\tau^-\to K^-K_S\nu_{\tau})$ is 2.3\%. 
For the $m_{ K^-K_S}$ spectrum this uncertainty varies  from 9\% at 
$m_{ K^-K_S}<1.1$ GeV/$c^2$ to 1\% at 1.7 GeV/$c^2$.

The 2\% uncertainty in the correction factor $p$ (Section \ref{rmspe}),
associated with $\tau$ branching fractions without a $\pi^0$,
leads to the 0.3\% uncertainty 
in the branching ratio. The mass-dependent uncertainty is
2\% at $K^-K_S$ mass below 1.1 GeV and 0.1\% for 1.7 GeV/$c^2$.

The systematic uncertainties from different sources, shown in
Table \ref{tab-sys}, are
combined in quadrature. The total systematic uncertainty for
the branching fraction ${\cal B}(\tau^-\to K^-K_S\nu_{\tau})$ is 2.7\%.
The systematic uncertainties for the mass spectrum are listed
in Table~\ref{tab-2}. They gradually decrease from $\simeq$9\% at 
$m_{ K^-K_S}=1$ GeV/$c^2$ to 1.5\% at $m_{ K^-K_S}=m_\tau$.
Near the maximum of the mass spectrum (1.3 GeV/$c^2$)
the uncertainty is about 2.5\%. 
\begin{figure}
\includegraphics[width=0.45\textwidth]{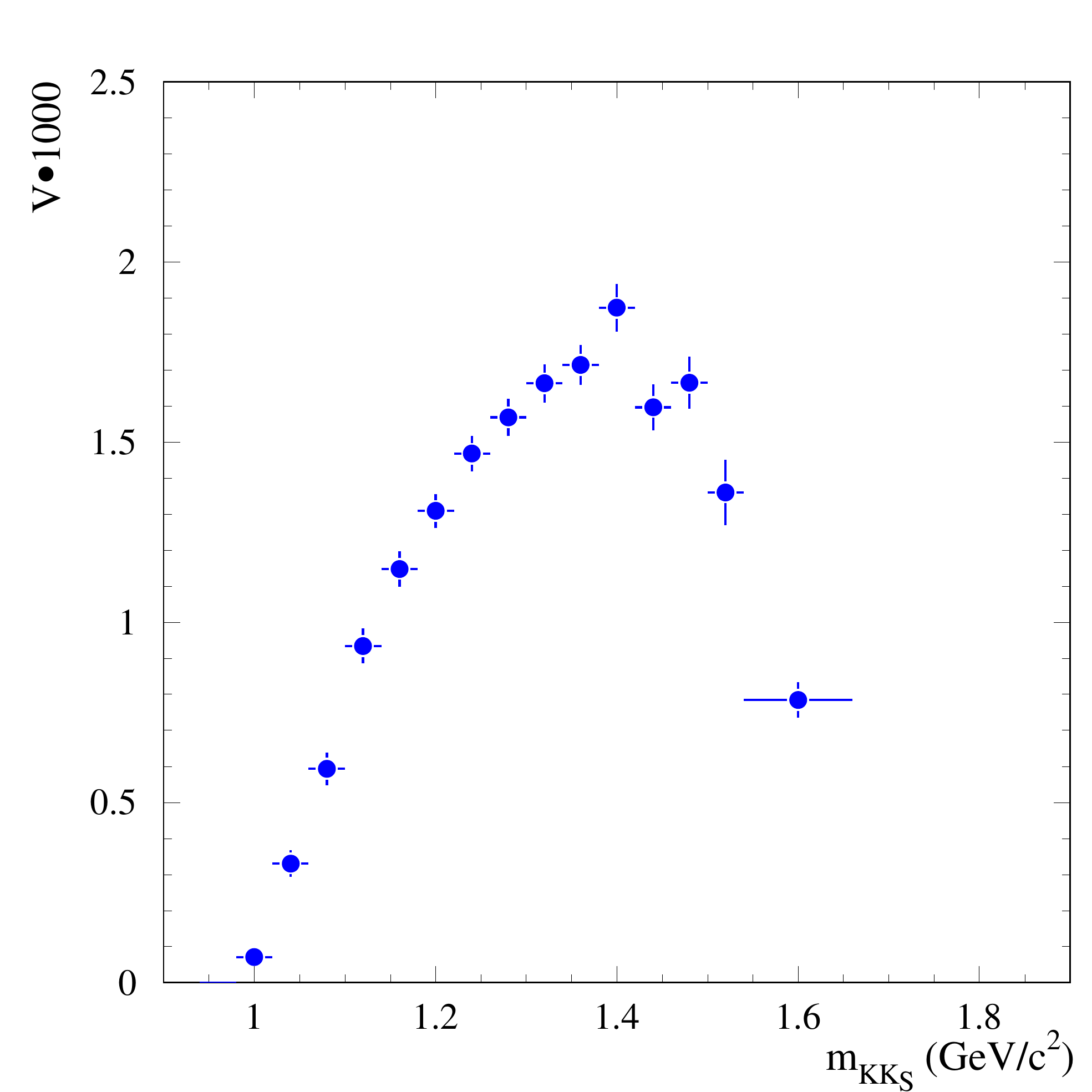} 
\caption { Measured spectral function for the 
$\tau^- \to K^-K_S\nu_\tau$ decay. Only statistical uncertainties are shown.
\label{vsfcs}}
\end{figure}

\begin{table}
\caption{
Measured spectral function (V) of the $\tau^- \to K^-K_S\nu_{\tau}$
decay,
in bins of $m_{ K^-K_S}$.
The columns report: the range of the bins, the normalized number of
events, the value of the spectral function. The first error is statistical, 
the second systematic. 
\label{tab-2}}
\begin{ruledtabular}
\begin{tabular}{ccc}
$m_{ K^-K_S}$(GeV/c$^2$)  &$N_s/N_{tot}\times 10^3$ &$V\times10^3$     \\ \hline
$0.98-1.02$  &$5.6 \pm1.4 $ &$0.071\pm0.018\pm0.006 $   \\  
$1.02-1.06$  &$26.0\pm2.7 $ &$0.331\pm0.034\pm0.026 $  \\ 
$1.06-1.10$  &$46.0\pm3.2 $ &$0.593\pm0.042\pm0.042 $   \\  
$1.10-1.14$  &$70.8\pm3.5 $ &$0.934\pm0.046\pm0.056 $  \\ 
$1.14-1.18$  &$84.4\pm3.4 $ &$1.148\pm0.047\pm0.057 $  \\  
$1.18-1.22$  &$92.3\pm3.3 $ &$1.309\pm0.046\pm0.052 $  \\ 
$1.22-1.26$  &$98.2\pm3.2 $ &$1.468\pm0.048\pm0.044 $ \\  
$1.26-1.30$  &$98.4\pm3.2 $ &$1.569\pm0.050\pm0.042 $  \\ 
$1.30-1.34$  &$96.3\pm3.0 $ &$1.663\pm0.052\pm0.042 $  \\  
$1.34-1.38$  &$90.2\pm2.9 $ &$1.715\pm0.052\pm0.039 $   \\ 
$1.38-1.42$  &$87.8\pm3.1 $ &$1.873\pm0.066\pm0.039 $  \\  
$1.42-1.46$  &$65.1\pm2.6 $ &$1.597\pm0.064\pm0.032 $  \\ 
$1.46-1.50$  &$57.3\pm2.5 $ &$1.666\pm0.073\pm0.032$    \\  
$1.50-1.54$  &$38.1\pm2.5 $ &$1.361\pm0.090\pm0.023 $  \\ 
$1.54-1.66$  &$36.9\pm2.4 $ &$0.785\pm0.049\pm0.013 $   \\  
$1.66-1.78$  &$6.6 \pm10.2$ &$0.986\pm1.520\pm0.014  $   \\ 
\end{tabular}
\end{ruledtabular}
\end{table}

\section{The results }
The branching ratio of the $\tau^-\to K^-K_S\nu_{\tau}$ decay is obtained
using the following expression:
\begin{eqnarray}
{\cal B}(\tau^{-}\to K^{-}K_S\nu_{\tau})&=& \frac{N_{\rm exp}}
{2 L B_{\rm lep} \sigma_{\tau\tau}}=\nonumber\\
(0.739\pm 0.011\pm 0.020)\times 10^{-3},
\label{eq11}
\end{eqnarray}
where $N_{\rm exp}=223741~\pm~3461$ (error is statistical) 
is the total number of signal events in the
spectrum in Fig.~\ref{mspec}, $L=468.0\pm2.5$ fb$^{-1}$ 
is the \babar\ integrated luminosity \cite{rlumi},
$\sigma_{\tau\tau}=0.919\pm0.003$ nb is the 
$e^+e^-\to \tau^+\tau^-$ cross section
at 10.58 GeV ~\cite{kk2f}
and $B_{\rm lep}$=0.3521~$\pm$~0.0006 is the world average 
sum of electronic and muonic branching fractions of the  $\tau$ lepton \cite{PDG}.
The first uncertainty  in (\ref{eq11}) is the statistical, 
the second is systematic.  Our result agrees well with
the Particle Data Group (PDG) value
$(0.740\pm 0.025)\times 10^{-3}$~\cite{PDG}, which is determined mainly
by the recent Belle measurement
$(0.740\pm 0.007\pm 0.027)\times 10^{-3}$~\cite{Belle}.

The measured mass spectrum $m_{ K^-K_S}$ for the $\tau^- \to K^-K_S\nu_{\tau}$ 
decay is shown in Fig.~\ref{mspec} and listed in Table~\ref{tab-2}. 
Our $m_{ K^-K_S}$ spectrum is compared with 
the CLEO measurement~\cite{CLEOt}. The \babar\ and
CLEO spectra are in good agreement.
The spectral function $V(q)$ calculated using 
Eq.~(\ref{eq1}) is shown in Fig.~\ref{vsfcs} and listed in Table \ref{tab-2}.
Due to the large error in the mass interval 1.66-1.78 GeV/c$^2$, which exceeds 
the scale of Fig.~\ref{vsfcs}, the value of $V(q)$ in this interval is not
shown in Fig.~\ref{vsfcs}.

\section{Conclusions}
The $K^-K_S$ mass spectrum and vector spectral function in the 
$\tau^-\to K^-K_S\nu_{\tau}$ decay have
been  measured by the \babar\ experiment. 
The measured $K^-K_S$ mass spectrum is far more precise than 
CLEO measurement \cite{CLEOt} and the branching fraction
$(0.739\pm 0.011\pm 0.020)\times 10^{-3}$ is comparable to 
Belle's measurement \cite{Belle}.

\section{ACKNOWLEDGMENTS}
We are grateful for the 
extraordinary contributions of our \pep2\ colleagues in
achieving the excellent luminosity and machine conditions
that have made this work possible.
The success of this project also relies critically on the 
expertise and dedication of the computing organizations that 
support \babar.
The collaborating institutions wish to thank 
SLAC for its support and the kind hospitality extended to them. 
This work is supported by the
US Department of Energy
and National Science Foundation, the
Natural Sciences and Engineering Research Council (Canada),
the Commissariat \`a l'Energie Atomique and
Institut National de Physique Nucl\'eaire et de Physique des Particules
(France), the
Bundesministerium f\"ur Bildung und Forschung and
Deutsche Forschungsgemeinschaft
(Germany), the
Istituto Nazionale di Fisica Nucleare (Italy),
the Foundation for Fundamental Research on Matter (The Netherlands),
the Research Council of Norway, the
Ministry of Education and Science of the Russian Federation, 
Ministerio de Econom\'{\i}a y Competitividad (Spain), the
Science and Technology Facilities Council (United Kingdom),
and the Binational Science Foundation (U.S.-Israel).
Individuals have received support from the Russian Foundation for Basic Research
(grant No. 16-02-00327),
the Marie-Curie IEF program (European Union) and the A. P. Sloan Foundation (USA).

\end{document}